\documentclass[pra,showpacs,twocolumn,superscriptaddress]{revtex4}

\newcommand{\ket}[1]{|{#1}\rangle}

\newcommand{\be}{\begin{equation}}
\newcommand{\ee}{\end{equation}}

\usepackage{amsfonts}
\usepackage{amsmath}
\usepackage{graphicx}
\usepackage{amssymb}
\usepackage{amsmath}
\usepackage{amssymb}
\usepackage{graphicx}
\usepackage{lscape}
\usepackage{color}
\usepackage{epstopdf}

\begin{document}
\title{Non-Galilean response of Rashba coupled Fermi gases}

\author{Daniel Maldonado-Mundo}
\affiliation{SUPA, Institute of Photonics and Quantum Sciences, Heriot-Watt University, Edinburgh EH14 4AS, United Kingdom}
\author{Lianyi He}
\affiliation{Frankfurt Institute for Advanced Studies and Institute for Theoretical Physics, J.~W. Goethe University, 60438 Frankfurt am Main, Germany}
\author{Patrik \"Ohberg}
\affiliation{SUPA, Institute of Photonics and Quantum Sciences, Heriot-Watt University, Edinburgh EH14 4AS, United Kingdom}
\author{Manuel Valiente}
\affiliation{SUPA, Institute of Photonics and Quantum Sciences, Heriot-Watt University, Edinburgh EH14 4AS, United Kingdom}
\begin{abstract}
We consider the effect of a momentum kick on the ground state of a non-interacting two-dimensional Fermi gas subject to Rashba spin-orbit coupling. Although the total momentum is a constant of motion, the gas does not obey the rules of Galilean relativity. Upon imprinting a small overall velocity to the non-interacting gas, we find that the Fermi sea is deformed in a non-trivial way. We also consider a weakly repulsive Fermi gas, and find, from its Hartree shift, that the total ground state of the system may change into a deformed, finite momentum ground state as the repulsion is increased beyond a critical value, without the need of any external Zeeman fields. We also discuss possible experimental signatures of these effects.  
 
\end{abstract}
\maketitle
\section{Introduction}
Recent progress on synthetic spin-orbit coupling for neutral atoms \cite{ref1,ref2,ref3,ref4} provides new ways to study its effects in both bosonic \cite{ref5,ref6,ref7} and fermionic \cite{ref8,ref9} systems. For atomic Fermi gases, many theoretical studies have shown that the spin-orbit coupling has nontrivial consequences on the BCS-BEC crossover when an attractive interaction is present \cite{ref10,ref11,ref12,ref13,ref14}. Even when the attraction is small, the BCS-BEC crossover can be induced by increasing the spin-orbit coupling strength. Due to the absence of Galilean invariance in the spin-orbit coupled gases, fermionic superfluids can exhibit highly non-trivial properties \cite{ref15,ref16,ref17}. These systems can undergo quantum phase transitions to some topological superfluid phases \cite{ref18,ref19,ref20,ref21}, and it has also been shown that the finite momentum pairing state or the so-called FFLO state can be energetically favored in spin-orbit coupled Fermi gases \cite{ref22,ref23,ref24,ref25,ref26}. However, these more exotic phenomena require an additional, carefully engineered Zeeman field. 

In this work, we study the response of a Fermi gas with pure Rashba spin-orbit coupling to a small momentum kick, in two spatial dimensions for concreteness. We find that, even in the low-momentum regime, the Fermi sea is deformed in a non-trivial way, is highly degenerate, and that its energy is bound from above by that of trivial Galilean-like boosts. We then consider repulsive interactions. {Unlike the bosonic case, where an energetic instability can be induced by a momentum kick \cite{Ozawa2},} we find, using mean-field theory, that the ground state acquires a finite momentum for interaction strengths beyond a critical value. The Fermi sea is consequently deformed, without the need to include any external Zeeman fields. We also describe realistic experimental signatures of these effects in systems of ultracold atoms with the use of standard techniques. 

\section{Non-interacting spin-orbit coupled Fermi gas}
The single-particle Hamiltonian of the system is given by \cite{Ozawa}
\begin{equation}
H_0 =\frac{\mathbf{p}^2+\hbar^2\lambda^2}{2m}\hat{1}+\frac{\hbar\lambda}{m}\mathbf{\sigma}\cdot \mathbf{p},\label{Ham}
\end{equation}
where $\mathbf{\sigma}=(\sigma_x,\sigma_y)$ is the vector of spin-$1/2$ Pauli matrices. The constants of motion are the helicity $\mathcal{H}\equiv \mathbf{\sigma}\cdot \mathbf{p}$,  with its corresponding eigenvectors $\ket{\psi^{(\pm)}}$ forming the helicity basis, with eigenvalues $h=\pm |\mathbf{p}|$ for momentum $\mathbf{p}=\hbar \mathbf{k}$, and the momentum itself. The common eigenstates of the Hamiltonian, helicity and momentum have the form
\begin{equation}
\ket{\psi^{(\pm)}(\mathbf{r})}=e^{i\mathbf{k}\cdot \mathbf{r}}\left[\ket{\uparrow}\pm e^{i\phi}\ket{\downarrow}\right],\label{singleparticlestates}
\end{equation} 
where $\phi$ is the polar angle of $\mathbf{k}$, given by $\tan{\phi}=k_y/k_x$. The two energy branches of the system, corresponding to negative (lower branch) and positive (upper branch) helicity, respectively, are given by
\begin{equation}
\epsilon_{\pm}(\mathbf{k})=\frac{\hbar^2}{2m}(|\mathbf{k}|\pm \lambda)^2.\label{singleparticleenergy}
\end{equation}

For the sake of simplicity, we will assume that the Fermi gas is dilute enough that in the ground state particles only occupy negative helicity states. For this to hold, the Fermi energy  must be $E_F\equiv \hbar^2k_F^2/2m \le \hbar^2\lambda^2/2m$, with the Fermi momentum $k_F=\pi\rho/\lambda$. Defining the only dimensionless parameter of the gas $z=\pi \rho/\lambda^2$, the single-branch condition reads $z<1$. The Fermi sea in two dimensions has the form of a concentric annulus. For convenience, we define ``radii'' $R_I=\lambda-k_F=\lambda(1-z)$ and $R_O=\lambda+k_F=\lambda(1+z)$ for the inner and outer circumferences, respectively, which are the borders of the Fermi sea. The ground state energy density $\mathcal{E}$ is given by
\begin{equation}
\mathcal{E}_0=\frac{1}{(2\pi)^2}\int_{0}^{2\pi}d\phi\int_{R_I}^{R_O}dk k \epsilon_{-}(\mathbf{k})= \frac{\hbar^2\rho}{6m}\pi z\rho.\label{noninteractingGS}
\end{equation}

We study now the properties of the lowest energy Fermi sea when the gas is given an infinitesimally small momentum kick. The momentum kick per particle is denoted as $\mathbf{k}_0$. Without loss of generality, we choose the momentum kick in the positive $x$-direction, $\mathbf{k_0}=k_0\hat{k}_x$, with $k_0>0$.

We begin by considering the obvious Galilean-like boost, consisting of assigning an extra momentum $k_0$ to each fermion in the Fermi sea. We note that, formally, these are not Galilean boosts since they also involve a non-trivial spin rotation, as is observed from Eq. (\ref{singleparticlestates}). Geometrically, these transformations correspond to displacing the Fermi sea to the right by an amount $k_0$. The excess energy density $\Delta \mathcal{E}_G$ is calculated by replacing $\mathbf{k}\to \mathbf{k}+\mathbf{k}_0$ in the single-particle dispersion, Eq. (\ref{singleparticleenergy}), as
\begin{equation}
\Delta \mathcal{E}_G=\frac{1}{(2\pi)^2}\int_{0}^{2\pi}d\phi\int_{R_I}^{R_O}dk k \epsilon_{-}(\mathbf{k}+\mathbf{k}_0)-\mathcal{E}_0.\label{energyGalileanboost1}
\end{equation}
Before performing the integration, we expand the modulus $|\mathbf{k}+\mathbf{k}_0|$ in the single-particle dispersion to second order in $k_0$, given by
\begin{equation} 
|\mathbf{k}+\mathbf{k}_0|=k\left(1+\frac{k_0}{k}\cos\phi+\frac{k_0^2}{2k^2}\sin^2\phi\right)+O(k_0^3).
\end{equation}
Inserting the expansion in Eq. (\ref{energyGalileanboost1}), we obtain the excess energy for Galilean-like boosts to $O(k_0^2)$,
\begin{equation}
\Delta \mathcal{E}_G = \frac{\hbar^2k_0^2}{4m}\rho.\label{energyGalileanboost}.
\end{equation} 
Below, we will see that the energy of these transformations is an upper bound for the ground state energy at finite momentum, and is only attained for a saturated lower branch ($z=1$).

If the momentum kick per particle $\mathbf{k}_0$ to the gas is small, the ground state in the thermodynamic limit can be modelled by infinitesimal transformations, which describe the response of the system to small momentum kicks. These must fulfill two conditions: (i) the density $\rho$ is preserved and (ii) the momentum per particle of the resulting Fermi sea equals $\mathbf{k}_0$. If we consider infinitesimally small momenta, we can safely rule out breaking the Fermi sea into disjoint pieces. We are then left with three possibilities, namely displacements of the inner and outer circumferences and multipolar deformations of these \footnote{Another density-preserving transformation corresponds to scaling of the radii $R_O$ and $R_I$ by factors $\alpha$ and $\beta(\alpha)$, respectively. However, these only increase the energy without changing the momentum per particle.} 


We now consider displacements of the inner and outer circumferences with respect to each other, which always preserve the density of the system. The resulting momentum density is easy to calculate if we regard the empty region $|\mathbf{k}|<R_I$ as a virtual Fermi sea of holes with density $\rho_I$, and the partially filled region $|\mathbf{k}|<R_O$ as a virtual Fermi sea of particles with density $\rho_O$, which are defined as
\begin{equation}
\rho_j=\frac{1}{(2\pi)^2}\int_{0}^{2\pi}d\phi\int_{0}^{R_j}dk k=\frac{R_j^2}{4\pi},\label{rhoj}
\end{equation}
where $j=I,O$.
If we displace the inner and outer circumferences by momenta $\mathbf{q}_I$ and $\mathbf{q}_O$, respectively, the momentum density of the system is given by
\begin{equation}
\mathbf{k}_0\rho=\mathbf{q}_O\rho_O-\mathbf{q}_I\rho_I.
\end{equation}
   
Finally, deformations of the Fermi sea can be parametrized in polar coordinates by allowing the radii become angle-dependent as $R_I(\phi)=R_I+f_I(\phi)$ and $R_O(\phi)=R_O+f_O(\phi)$, with $f_I$ and $f_O$ real periodic functions. Clearly, not all deformations preserve the density. This is given by $\rho=\rho_O-\rho_I$, which is computed from Eq. \ref{rhoj} with $R_j\to R_j(\phi)$. This implies the following condition
\begin{equation}
\int_{0}^{2\pi}d\phi \left\{\left[f_O(\phi)\right]^2-\left[f_I(\phi)\right]^2\right\}=0,\label{norms}
\end{equation}
or, equivalently $||f_O||=||f_I||$, with $||\cdot||$ the norm on $L^2([0,2\pi))$. This condition is very relaxed, since it allows very different deformations of the two circumferences provided that their norms are equal. Since, by convention, we have chosen to imprint a momentum to the system in the positive $x$-direction, we can restrict ourselves to deformations in the horizontal axis. Therefore, we parametrize the functions $f_i$ ($i=O,I$) by a multipolar expansion of the form
\begin{equation}
f_i(\phi)=\sum_{m\ge 1} c_m^{(i)}\frac{\cos(m\phi)}{\sqrt{\pi}},\label{multipolar}
\end{equation}  
where every coefficient $c_m^{(i)}$ is real. The condition on the norms takes the form $\sum_{m\ge 1}(c_m^{(O)})^2=\sum_{m\ge 1}(c_m^{(I)})^2$. The momentum density of the deformed Fermi sea is given by $\mathbf{k_0}\rho=k_0\rho\hat{k}_x$, with
\begin{equation}
k_0\rho = \frac{1}{(2\pi)^2}\int_{0}^{2\pi}d\phi\int_{R_I+f_I(\phi)}^{R_O+f_O(\phi)}dk k^2\cos{\phi}.
\end{equation}
For an infinitesimally small momentum kick, we can restrict the expansion in Eq. (\ref{multipolar}) to lowest (dipolar) order, in which case condition (\ref{norms}) reduces to $c_1^{(I)}=\pm c_1^{(O)}$. Defining $q_1^{\pm}=c_1^{(O)}/\sqrt{\pi}$, with the upper (lower) sign corresponding to equal (opposite) dipolar coefficients, we have $k_0= q_1^+$, while for $c_1^{(I)}=-c_1^{(O)}$ we have
\begin{equation}
k_0\rho = \left(\rho+\frac{R_I^2}{2\pi}\right)q_1^{-}+\frac{(q_1^{-})^3}{8\pi}.\label{equationopposite}
\end{equation}
To lowest order in $k_0$, $q_1^{-}$ is given by $q_1^{-}=2zk_0/(1+z^2)$. Intuitively, dipolar coefficients with opposite sign will yield a lower energy, since a weaker deformation ($|q_1^{-}|<|q_1^+|$) is needed in order to have the required momentum $k_0$. The excess energy density $\Delta \mathcal{E}_{\pm}$ for dipolar deformations is readily calculated, to leading order in $q_1^{\pm}$, giving
\begin{equation}
\Delta \mathcal{E}_{\pm} = \frac{\hbar^2(q_1^{\pm})^2}{4m}\rho.
\end{equation}
From the above relation, we see that for equal deformations the excess energy is that of Galilean-like boosts, as in Eq. (\ref{energyGalileanboost}), $\Delta \mathcal{E}_+=\Delta \mathcal{E}_G$. For opposite dipolar coefficients we obtain a lower energy given by
\begin{equation}
\Delta \mathcal{E}_{-} = \frac{\hbar^2k_0^2}{m}\left[\frac{z}{1+z^2}\right]^2\rho,\label{energydeformations}
\end{equation}
which is bound from above by $\Delta \mathcal{E}_G$.  The excess energy due to pure deformations can be lowered further by considering either displacements or both displacements and deformations. We will show next that the resulting ground-state at small non-zero momentum is highly degenerate. 

We now study the energy of the Fermi gas when we impose displacements $\mathbf{q}_I$ and $\mathbf{q}_O$ of the inner and outer circumferences, together with deformations of opposite coefficients, parametrized by $\mathbf{q}_1\equiv \mathbf{q}_1^{-}$. The energy density of the system is calculated as the difference between the energy of the virtual Fermi sea of particles, $\mathcal{E}_O$, and that of the virtual Fermi sea of holes, $\mathcal{E}_I$, where
\begin{equation}
\mathcal{E}_i=\frac{1}{(2\pi)^2}\int_{0}^{2\pi}d\phi \int_{0}^{R_i+f_i(\phi)}dk k \epsilon_{-}(\mathbf{k}+\mathbf{q}_i),
\end{equation}
with $i=O,I$. To lowest order in the momenta, we can isolate the excess energy from the energy at zero momentum, which takes the form
\begin{align} 
 (2\pi)^2\frac{2m}{\hbar^2}\Delta \mathcal{E} &= \pi k_F^2(q_O^2-q_I^2)+\pi\lambda k_F(q_O^2+q_I^2)\nonumber \\
&+2\pi k_F\left[k_Fq_1(q_O+q_I)+\lambda q_1(q_O-q_I)\right]\nonumber \\
&+2\pi\lambda k_Fq_1^2.\label{excessenergygeneral}
\end{align}
If the momentum per particle is set to $\mathbf{k}_0=k_0\hat{k}_x$, the three momenta of the system must fulfill the condition
\begin{equation}
q_O\rho_O-q_I\rho_I+\frac{1+z^2}{2z}\rho q_1=k_0\rho.\label{condition}
\end{equation}
Using the momentum constrain above to clear $q_I$ in favor of $q_O$ and $q_1$, and inserting this in to Eq. (\ref{excessenergygeneral}), we obtain an expression for the excess energy density, which is minimized for any $q_O$ satisfying
\begin{equation}
q_O=-q_1+\frac{2z(1+z)}{1+3z^2}k_0,
\end{equation}
with the only constrain that both $q_O$ and $q_1$ have to be small ($\ll k_F$) for our theory to be valid. We thus see that the ground-state at fixed finite momenta is infinitely degenerate. Finally, the excess energy density is given by
\begin{equation}
\Delta \mathcal{E} = \frac{\hbar^2k_0^2}{m}\frac{z^2}{1+3z^2}\rho,\label{excessenergydisp}
\end{equation} 
which is bound from above by $\Delta \mathcal{E}_G$ for all $z$ and by $\Delta \mathcal{E}_{-}$ for each $z$.

\section{Weakly-interacting Fermi gas}
We consider now the effect of weak interactions, which we model with a zero-range potential. At mean-field level, the Dirac delta needs not be regularized, and has the form $g\delta^{(2)}(\mathbf{r})$, where $\mathbf{r}$ is the interparticle distance. The dimensionless interaction parameter $\xi=mg/4\pi\hbar^2$ \cite{Petrov} must here be compared with the only dimensionless parameter $z$ of the non-interacting gas. The typical interaction energy $g\rho/4$ \cite{Maldonado2013} should be smaller than the Fermi energy $E_F=\pi\hbar^2\rho z/2m$, which gives $\xi< z/2$. 

The Hartree shift $\mathcal{E}^{(1)}$ is given by \cite{Maldonado2013}
\begin{equation}
\mathcal{E}^{(1)}=\frac{g}{4}\rho^2-\frac{g}{4(2\pi)^4}\int_Fd\mathbf{k}\int_Fd\mathbf{k}'\cos(\phi_{\mathbf{k}}-\phi_{\mathbf{k}'}).
\end{equation}
The first term is independent of the shape of the Fermi sea, and the second term vanishes identically in the non-interacting ground state. However, the second term reduces the interaction energy for repulsive interactions ($g>0$) if the Fermi sea is modified to acquire a non-zero total momentum. Without loss of generality, we again consider a momentum per particle $\mathbf{k}_0$ in the $x$-direction, in which case the extra contribution to the Hartree shift reads 
\begin{align}
\mathcal{I}&\equiv-\frac{g}{4(2\pi)^4}\int_Fd\mathbf{k}\int_Fd\mathbf{k}'\cos(\phi_{\mathbf{k}}-\phi_{\mathbf{k}'})\nonumber \\
&= -\frac{g}{4}\left[\frac{1}{(2\pi)^2}\int_Fd\mathbf{k}\cos\phi\right]^2.\label{extraHartree}
\end{align}

In the following, we minimize the total energy excess $\Delta \mathcal{E}_T = \Delta \mathcal{E}+\mathcal{I}$ at a constant, infinitesimaly small momentum $\mathbf{k}_0$, with respect to the displacement parameter $\mathbf{q}_O$. We consider displacements, parametrized by $q_I$ and $q_O$, together with deformations, parametrized by $q_1\equiv q_1^-$. The Fermi sea is the set $F$ of momenta $\mathbf{k}$ defined as $F=\{\mathbf{k}|\kappa_I\le |\mathbf{k}|\le \kappa_O\}$, with $\kappa_i(\phi) = \sqrt{(R_i\pm q_1\cos\phi)^2+q_i^2+2q_i(R_i\pm q_1\cos\phi)\cos\phi}$ ($i=I,O$).
The extra Hartree shift of Eq. (\ref{extraHartree}) is easily calculated and reads
\begin{equation}
\mathcal{I}=-\frac{g\rho}{64\pi z}\left[q_O (1+z) - q_I (1-z) + 2q_1\right]^2,\label{previousHartree}
\end{equation}
where $q_O$ and $q_I$ are related via Eq. (\ref{condition}), which is now valid to lowest order in $k_0$. The energy excess $\Delta \mathcal{E}_T$ is minimized at the point 
\begin{equation}
q_{O}  = \frac{2k_0 z (-2+2\xi[1-z])+q_1(2-z-2z[z+\xi])}{2(-1+2z^2+z\xi[1-z^2])},\label{point}
\end{equation}
and its value is
\begin{equation}
\Delta \mathcal{E}_T = \mathcal{F}(\xi,z)\frac{\hbar^2k_0^2}{m}\rho,\label{finalhartree}
\end{equation} 
where
\begin{equation}
\mathcal{F}(\xi,z)=z\frac{z-\xi}{\xi z(z^2-1)+1+3z^2}.\label{F}
\end{equation}
Note that $\mathcal{F}(\xi,z)$ is independent of choice of deformation parameter $q_1$, which shows the infinite degeneracy at small momenta holds even in the presence of interactions.
As we infer from Eq. (\ref{finalhartree}), the system's ground state, to first order, will change from the non-interacting Fermi sea to a Fermi sea with an infinitesimally small momentum -- which denotes continuity -- at a critical value $\xi=\xi_c(z)$ where $\mathcal{F}(\xi_c,z)=0$. This is given by
\begin{equation}
\xi_c(z)=z.
\end{equation}
It is important to note that, while we are considering the system's non-interacting energy to order $O(k_0^2)$, the interaction energy is not approximate but is indeed proportional to $k_0^2$ in the cases $q_I=q_O=0$ with $q_1\ne 0$ and $q_1=0$ with $q_I,q_O\ne 0$, and therefore the above critical value is exact at the mean-field level. The critical value we have obtained is beyond the mean-field regime. This is a typical situation that also occurs, for instance, in the study of the ferromagnetic transition \cite{Stoner} in repulsive Fermi gases \cite{Huang,Kanno,Duine}, and is resolved by going beyond first-order perturbation theory, which can result in an apparent first-order transition \cite{Duine,Conduit}, or using non-perturbative methods \cite{HeHuang,Heiselberg}, which predict a second-order phase transition and are in good agreement with Monte-Carlo simulations \cite{Pilati}. Our findings are the natural starting point for higher-order corrections \cite{Maldonado2013}, non-perturbative treatments, {and can still be improved at the mean-field level \cite{Kinnunen}}. Moreover, if, as our results suggest, the transition to finite momentum is continuous, the critical point can be obtained by calculating only the second-order response function $\mathcal{F}(\xi,z)$ but with improved treatments of the interactions.   

The low-momentum theory we have presented only predicts whether the system evolves towards a finite momentum ground state in a continuous manner. This is to say that our theory can describe, correctly, only derivatives in the energy at zero momenta
\begin{equation}
\frac{m}{2\hbar^2\rho}\frac{d^2\mathcal{E}}{dk_0^2}(k_0=0)=\mathcal{F}(\xi,z).
\end{equation}
The first derivative of the energy with respect to $k_0$ vanishes always at $k_0=0$, while its second derivative $\mathcal{F}(\xi,z)>0$ for $\xi<\xi_c$ and $\mathcal{F}(\xi,z)\le 0$ for $\xi\ge \xi_c$ (strictly speaking, until its pole at $\xi_{\infty}=(1+3z^2)/(z-z^3)\gg \xi_c$, which is far beyond the limit of validity of the mean-field theory). The change of sign in $\mathcal{F}(\xi,z)$ denotes the transition from a minimum to a maximum at $k_0=0$, which implies that for repulsions stronger than the critical value $\xi_c$ the system's ground state has non-zero momentum. To calculate the actual momentum of the ground state for $\xi>\xi_c$ it is necessary to go beyond leading order in $k_0$. 

\section{Experimental considerations}
So far, we have considered the momentum kick of the gas to be given in a particular direction. In an experiment, however, no direction is in principle preferred. Obviously, after the critical point is reached, the ground-state is highly degenerate, since the same momentum in any direction yields the same energy. The many-body wave function is therefore an arbitrary superposition of states with momenta pointing at different directions. For instance, we may expect to observe an equal superposition of these states, up to arbitrary phases, which yields a circularly symmetric momentum distribution. Momentum distributions are especially relevant to cold atom experiments where these can be obtained via time-of-flight measurements \cite{Bloch}, in combination with spin-injection spectroscopy \cite{X} or momentum resolved RF spectroscopy \cite{Y}. 

\begin{figure}
\includegraphics[scale=0.3]{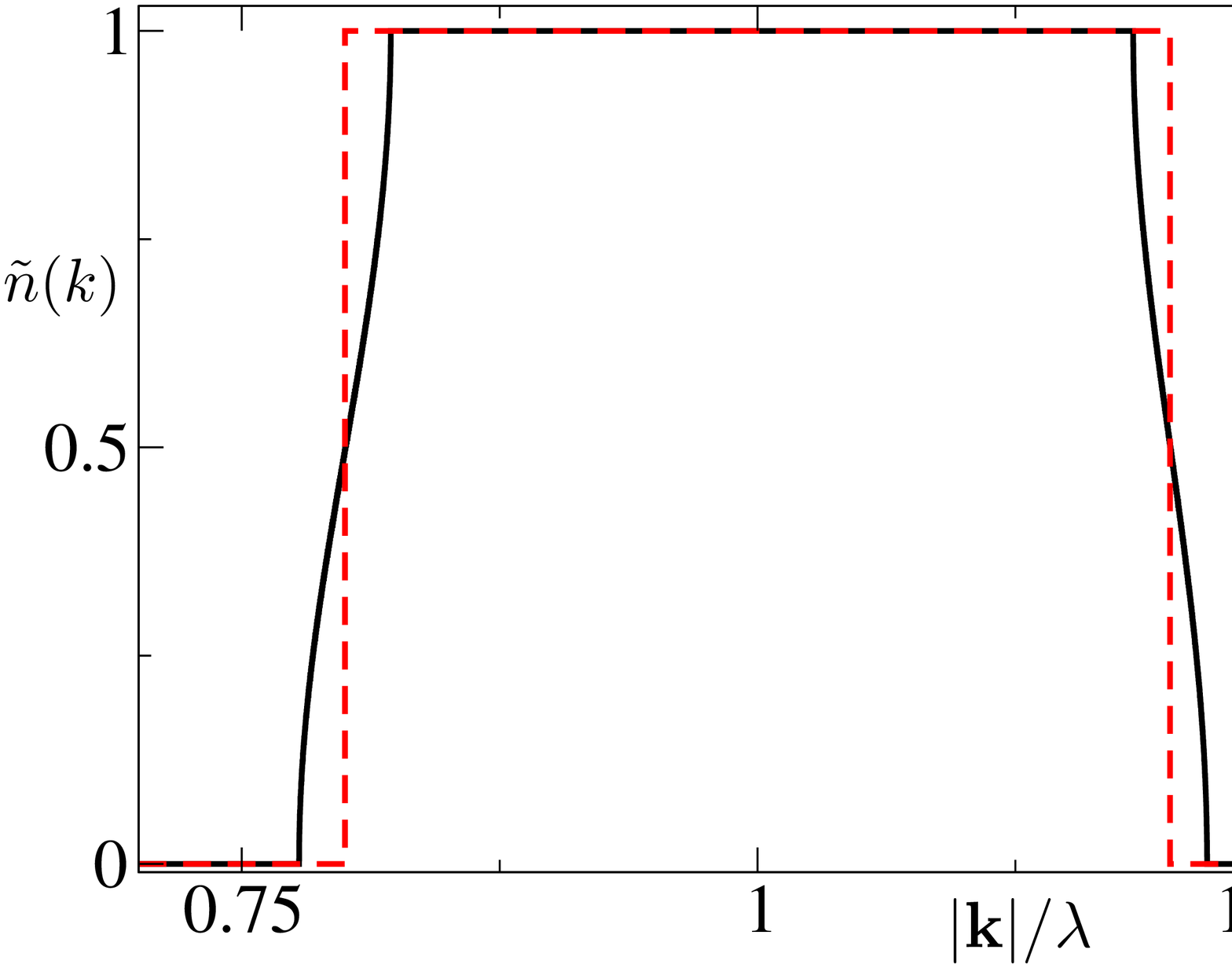}
\caption{\label{ref} (Color online) Integrated momentum distribution (black solid line) at critical interaction strength $\xi=z$, with $k_0/\lambda=5\cdot 10^{-2}$ and $z=1/5$, compared to the non-interacting momentum distribution (red dashed line).}
\end{figure}

We can easily map out, starting from a Fermi sea at finite momentum in a particular direction, the aforementioned circularly symmetric momentum distribution. The momentum distribution is denoted by $n(\mathbf{k})$, and is such that $\rho=\int d\mathbf{k}n(\mathbf{k})/(2\pi)^2$. We define the integrated, angle-independent momentum distribution $\tilde{n}(k)$ as
\begin{equation}
2\pi\tilde{n}(k)=\int_{0}^{2\pi}d\phi n(\mathbf{k}).
\end{equation}
Geometrically, $\tilde{n}(k)$ is the arc length of occupied states on a circumference of radius $k$ in momentum space. Closed circumferences map into unit length -- fully occupied states -- while open arcs give shorter lengths -- smaller average occupations. The circularly symmetric momentum distribution arising from the superposition of the different states is obtained as a surface of revolution by rotating the Fermi sea at finite momentum in a particular  direction with respect to the origin in momentum space. Clearly, the resulting momentum distribution coincides with $\tilde{n}(k)$. 

In Fig. \ref{ref}, we show the integrated momentum distribution for a particular case of an interacting Fermi sea at non-zero total momentum compared to the non-interacting momentum distribution. There, we observe a shortening of the unit occupation plateau, together with a smoothening of the momentum distribution at the edges of the Fermi sea, with an obvious change in concavity, which can be a relevant experimental signature for finite momentum states. In cold atom experiments, where an external trap is always present, the Fermi sea and the homogeneous momentum distribution can be observed by selectively probing fermions around the centre of the trap \cite{FermiSurface,Jin2}. An alternative way to observe these effects consists of adding to the system a small symmetry-breaking term, i.e. a small momentum kick in a chosen direction, in order to observe the deformations {\it per se}. This can routinely be done nowadays with the use of standing-wave light-pulse sequences \cite{Prentiss}. This technique has been successfully applied to ultracold atom systems subject to artificial spin-orbit coupling \cite{IanWaves}.

\section{Conclusions}
In this paper, we have studied the response of a dilute two-dimensional Fermi gas with Rashba spin-orbit coupling to a small overall constant velocity kick. We have found that the moving Fermi sea deforms in a non-trivial manner due to the non-Galilean nature of the system, and is highly degenerate. We have then considered repulsive interactions at the Hartree-Fock level, and found that the ground-state of the system acquires a finite-momentum. The Fermi sea becomes deformed beyond a critical interaction strength in a continuous fashion, which we identified as a possible experimental signature. Our results open the path towards the observation of finite momentum ground states, constitute the starting point for more elaborate treatments of interactions, and can be generalized to higher dimensions and more general types of spin-orbit coupling.    

\acknowledgements{D.M.-M. acknowledges support from the EPSRC CM-DTC, L.H. is supported by the Helmholtz International Center for FAIR within the
framework of the LOEWE program, P.\"O and M.V.  acknowledge support from EPSRC grant No. EP/J001392/1.}

\bibliographystyle{unsrt}

\end{document}